\documentstyle[floats,aps]{revtex}
\advance\topmargin 30pt
\begin{document}
\input psfig
\pssilent
\title{Evolving the Bowen--York initial data for spinning black holes}

\author{Reinaldo J. Gleiser$^1$, Carlos O. Nicasio$^{1,2}$,
Richard H. Price$^{3}$, Jorge Pullin$^2$}
\address{1. Facultad de Matem\'atica, Astronom\'{\i}a y F\'{\i}sica,
Universidad Nacional de C\'ordoba,\\ Ciudad
Universitaria, 5000 C\'ordoba, Argentina.}
\address{
2. Center for Gravitational Physics and Geometry, Department of
Physics,\\
The Pennsylvania State University, 
104 Davey Lab, University Park, PA 16802}
\address{
3. Department of Physics, University of Utah, Salt Lake City, Utah
84112.}
\date{Oct 11th, 1997}
\maketitle
\begin{abstract}
  The Bowen-York initial value data typically used in numerical
  relativity to represent spinning black hole are not those of a
  constant-time slice of the Kerr spacetime.  If Bowen-York initial
  data are used for each black hole in a collision, the emitted
  radiation will be partially due to the ``relaxation'' of the
  individual holes to Kerr form.  We compute this radiation by
  treating the geometry for a single hole as a perturbation of a
  Schwarzschild black hole, and by using second order perturbation
  theory. We discuss the extent to which Bowen-York data can be
  expected accurately to represent Kerr holes.
\end{abstract}

\pacs{4.30+x}

\vspace{-11cm} 
\begin{flushright}
\baselineskip=15pt
CGPG-97/10-2  \\
gr-qc/9710096\\
\end{flushright}
\vspace{9cm}

\section{Introduction}\label{sec:intro}

The description of the collision of two black holes, including the
total energy radiated and the waveforms to be measured by observers far
from the collision region, is at this time one of the most active
fields of research in general relativity. Since the latter theory has
a well posed initial value problem, a good deal of effort has been
devoted to finding solutions for the initial value problem, in the
form of initial data sets, that may represent slices of recognizable
physical processes involving black holes. One of the first examples of
this kind was given by Misner \cite{Mi}, who derived an initial data
set  representing the moment of time symmetry in the
head-on collision of two equal mass black holes, placed at an
arbitrary distance from each other. Further developments along this
line have provided a fair number of interesting initial data sets
\cite{BoYo,Cook}, that can be interpreted as representing isolated but
boosted and/or rotating black holes, or collisions involving two or 
more black holes.

Once one has the initial data, the next task is to study the evolution
and the consequent emission of gravitational waves arising from the
collision.  Due to the complexity of the evolution equations of
general relativity, numerical solutions of the full Einstein equations
are available at present only for equal mass, nonspinning, holes
undergoing a head-on (i.e., zero impact parameter) collision.
\cite{etaletal,boost}.

It has also been observed that for a restricted set of parameters, the
evolution can be well described by treating the system as a
perturbation of a single black hole \cite{PrPu}. This method, called
the ``close approximation,'' has produced results that are in
remarkable agreement with the full numerical results in the case where
the latter are available.  An appealing feature of the perturbation
method is the explicit control over the parameters characterizing the
perturbation. This, together with the development of some form of
``error bars,'' as in \cite{GlNiPrPucqg,GlNiPrPuprl}, can make the
method an important tool to make predictions of physical processes or
to provide comparison cases to test the reliability of numerical
methods.

The initial data set that is usually considered \cite{BoYo,Cook} in
the study of black hole collisions is generated using the conformal
approach, in which one assumes that the spatial metric is conformally
flat, and the maximal slicing condition is chosen for the extrinsic
curvature.  In the flat space, it is relatively simple to construct a
conformally related extrinsic curvature which guarantees that the
momentum constraint is solved. The Hamiltonian constraint is
subsequently solved, either numerically \cite{Cook} or via
approximations \cite{boost,PuCa}. In this construction there is a
series of simplifying assumptions and there is no claim that a
``generic'' solution has been found.  It is therefore not clear that
the particular solution generated is a true representation of the
physical problem in which one is interested.  An example of this is
the Bowen and York (BY)\cite{BoYo} solution for a single spinning
hole. It is known that this initial data set represents a dynamical
situation that evolves to a Kerr black hole asymptotically in the
future. Initially, however, the spacetime is not a Kerr spacetime, but
can be thought (somewhat inappropriately) to differ from a Kerr
solution in that it has some ``gravitational wave content,'' the
waves that will be radiated as the spacetime evolves towards Kerr.

One can argue that this ``gravitational wave content'' will be
radiated in a short time, and that the initial data will evolve
rapidly to a stationary Kerr black hole configuration, and therefore
will not greatly affect the radiation produced in a black hole
collision, as long as the holes are released far from each other. The
question is important enough to deserve a more careful answer. In
addition, numerical relativity codes cannot be accurately run for long
evolution times, so initial data will have to be specified at fairly
late times, with holes fairly close together, and with the possibility
that the ``gravitational wave content'' of the initial data is a major
part of the outgoing radiation.

The purpose of this paper is to analyze the radiation from the
individual holes. More specifically, we use the theory of
perturbations of the Schwarzschild spacetime. We consider that we have a
family of spacetimes depending on the parameter $\epsilon$, and that,
in appropriate coordinate systems, metrics of the family can be
expanded as 
\begin{equation}
g_{\mu \nu} = g^{(0)}_{\mu
\nu} + \epsilon h^{(1)}_{\mu \nu}
+ \epsilon^2 h^{(2)}_{\mu \nu}+\cdots \ .
\end{equation}
Here $g^{(0)}_{\mu \nu}$ the Schwarzschild metric, $h^{(1)}_{\mu \nu}$
is called the first order perturbation, and $h^{(2)}_{\mu \nu}$ is
called the second order perturbation.  To analyze the ``Bowen and York
spacetime,'' the spacetime that evolves from Bowen and York initial
data for a single spinning hole, we choose as the expansion parameter
the angular momentum $J$, and we analyze both the Kerr solution and
the BY spacetime to second order in $J$. The second order expansions
are then compared and we find that we need only to evolve the
difference between the BY spacetime and the Kerr spacetime.

The organization of this paper is as follows.  In section II we give a
perturbation analysis of the BY initial data.  The method of evolving
this initial data is described in section III.  In section IV the
results for radiation emitted are presented and discussed.  In an
Appendix we show how the Kerr metric can be written as an expansion,
in angular momentum, about the Schwarzschild spacetime.

When it is useful to specify orders of expansion, and/or multipole
indices, we shall use a leading subscript to denote the multipole
index $\ell$, and a superscript, in parenthesis, to the right of a
perturbation variable, to indicate order in $J$. The quantity
$_0\psi^{(2)}$, for example, is a monopole perturbation, second order
in $J$.

%%%%%%%%%%%%%%%%
\section{The Bowen--York single rotating black hole}\label{sec:by}
%%%%%%%%%

The Bowen--York \cite{BoYo} construction of initial data assumes that
space-time contains a (constant $t$) slice, where the 3-metric can be
written in the form $g_{ij} = \psi^4 f_{ij}$, where $f_{ij}$ is 
the metric for the flat background
$ds^2=dR^2+R^2(d\theta^2+\sin^2{\theta}\,d\phi^2)$,
while the extrinsic curvature is given by $K_{ij} = \psi^{-2}
\widehat{K}_{ij}$, and satisfies $K^i_{i}=0$. With these assumptions
the initial value constraint equations take the form
\begin{equation}
\widehat{K}^{ij}_{\ \ |j}=0, 
\end{equation}
and
\begin{equation} 
\nabla^2 \psi = - {1 \over 8} \widehat{K}_{ij}  \widehat{K}^{ij} \psi^{-7},
\label{BaY} 
\end{equation}
where the Laplacian, and the covariant differentiation (denoted by a
vertical bar) are taken with respect to the flat space metric. The
problem is not fully specified until appropriate boundary conditions are
imposed on $\psi$. The Bowen--York prescription for a single
black hole is that, given a
certain constant $a$, Eq.\ (\ref{BaY}) holds for $R \geq a$, and
\begin{equation} 
{\partial \psi \over \partial R} +{1 \over 2R} \psi =0 
\;\; \mbox{for} \;\; R= a,   
\end{equation}
and
\begin{equation} 
  \psi > 0 
\;\; , \;\; \lim_{R \rightarrow \infty} \psi = 1.   
\end{equation}

The particular solution we are interested in corresponds to \cite{BoYo}
\begin{equation}\label{BYhatK}
\widehat{\bf{K}} = {3 \over R^3} \left[ (\bf{J}\wedge \bf{R}) \otimes \bf{R} 
+ \bf{R}\otimes (\bf{J}\wedge \bf{R})  \right],
\end{equation}
where $\bf{R}$ is the ``position'' vector in the flat space
background, and $\bf{J}$ is a vector constant. Without loss of
generality, we choose ${\bf J}= J {\bf k}$, where ${\bf k}$ is a unit
vector pointing along (the positive direction of)
the polar axis of the
$R,\theta,\phi$ coordinate system. With this choice, the only nonvanishing
components of $\widehat{\bf{K}}$ are
\begin{equation}\label{BYK}  
\widehat{K}_{R \phi} =\widehat{ K}_{\phi R} =  
{3 J \over R^2} \sin^2{\theta}\ ,
\end{equation} 
We then find
\begin{equation} 
\widehat{K}_{ij}  \widehat{K}^{ij} =  18 {J^2 \over R^6} \sin^2{\theta}\ .
\end{equation} 

We now need to solve Eq.\ (\ref{BaY}). For general values of $J$, this can
only be done numerically. However, we are really interested in
solutions near $J=0$ (``slow rotation'') and an expansion in powers
of $J$ is appropriate. The zeroth order equation is
\begin{equation} 
\nabla^2 \psi = 0\ ,
\label{BaY0e} 
\end{equation}
and the solution that satisfies the boundary conditions is
\begin{equation} 
 \psi^{(0)} = 1 + a/R \ .
\label{BaY0s} 
\end{equation}

The lowest order correction to $\psi$ may now be constructed by
linearizing $\psi$ about $\psi^{(0)}$ in Eq.\ (\ref{BaY}). If we
formally write
\begin{equation} 
 \psi = \psi^{(0)} + J^2 \psi^{(2)} + \dots\ ,
\end{equation}
the resulting equation is
\begin{equation} 
\nabla^2 \psi^{(2)} = -   {9 \over 4 R^6} \sin^2{\theta}
\left(1 + {a \over R}\right)^{-7}\ ,
 \label{BaY1e} 
\end{equation}
and the $\sin^2{\theta}$ factor may be expanded in Legendre polynomials,
as $2[P_0(\cos\theta)-P_2(\cos\theta)]/3$.

We next  write $\psi^{(2)}(R,\theta)$ as
\begin{equation} 
 \psi^{(2)}(R,\theta) =\  _0\psi^{(2)}(R)\,P_0(\cos{\theta}) 
+\ _2\psi^{(2)}(R)\ P_2(\cos{\theta})
\end{equation}
and we find that $_0\psi^{(2)}(R)$ and  $_2\psi^{(2)}(R)$ satisfy the
equations
\begin{equation} 
 {d^2\left[ _0\psi^{(2)}\right] \over dR^2} +{2 \over R}{d\left[
_0\psi^{(2)}\right]\over dR} = - {3 \over 2 R^6} \left(1 + {a \over
R}\right)^{-7} \ ,
\end{equation}
and
\begin{equation} 
 {d^2\left[ _2\psi^{(2)}\right] \over dR^2} +{2 \over R}{d\left[
_2\psi^{(2)}\right] \over dR} -{6 \over R^2}\ _2\psi^{(2)} = {3 \over
2 R^6} \left(1 + {a \over R}\right)^{-7} \ .
\end{equation}
The solutions of these equations, satisfying the boundary conditions are
\begin{equation}\label{psi02} 
  _0\psi^{(2)}  ={a^4+R^4+5aR(R+a)^2 \over 40 a^3 (R+a)^5} 
\end{equation}
and
\begin{equation} 
  _2\psi^{(2)}  = - {R^2 \over 20 a (R+a)^5}\ . 
\end{equation}

\begin{figure}%fig1
\hbox{\psfig{figure=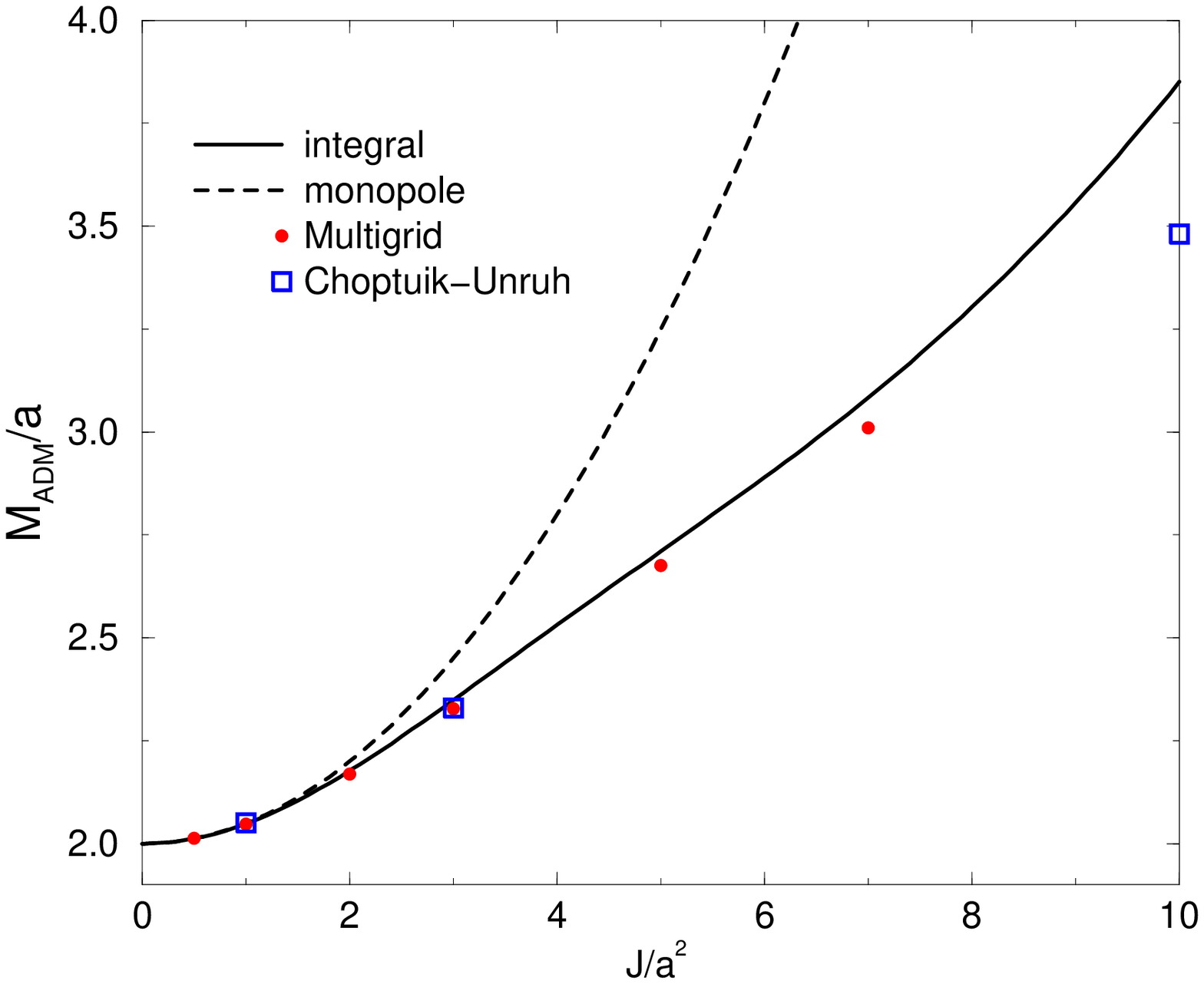,height=3in}
\psfig{figure=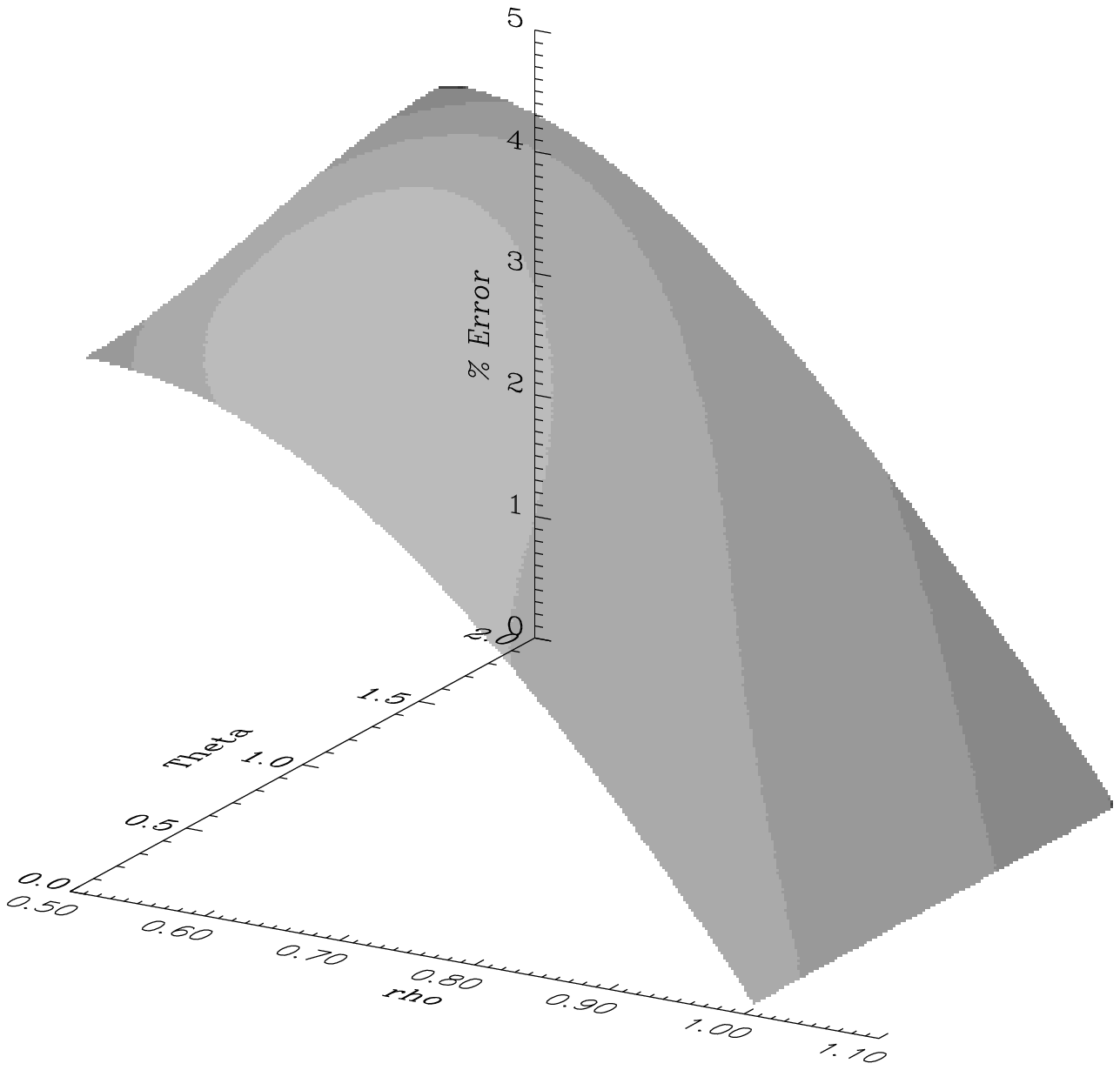,height=3in}}
\caption{Comparison of approximate solutions to the initial value
  problem with a full numerical integration performed with a multigrid
  method. Our code uses a square grid in compactified coordinates
  $\rho = R/(a+R)$, $\theta$ with $200\times 200$ grid points. It
  achieves an accuracy similar to the test runs of Choptuik and
  Unruh\protect\cite{ChUn}.  The figure at the left compares the ADM
  mass of approximate and numerical solutions. The two approximations are
  described in the text. The
  figure on the right shows the percentile difference between the full
  numerical solution for $\psi$ and the second order approximation,
  for $J/a^2=4$, as a function of $\rho,\theta$. We see that the
  approximation works very well.}
\label{fig:phicompar}
\end{figure}

With Eq.\ (\ref{psi02}) we see that the conformal factor has the form
\begin{equation} 
\psi=1+\frac{a}{R}+\frac{J^2}{40a^3}\,\frac{1}{R}
+{\cal O}(1/R^2)+{\cal O}(J^3)\ .
\end{equation}
But by the definition of the ADM mass (see for instance \cite{BoYo}),
$\psi=1+M_{\rm ADM}/(2R)+{\cal O}(1/R^2)$, and hence the ADM mass,
to second order in $J$, is
\begin{equation}\label{mADMbypert} 
M_{\rm ADM}=2a+\frac{J^2}{20a^3}\ .
\end{equation}
This dependence of $M_{\rm ADM}$ on $J$, for constant $a$, is plotted
in Fig.\,\ref{fig:phicompar} as the dashed curve. Two other
computations of $M_{\rm ADM}$ are also plotted, based on the integral
\cite{BoYo,ChUn},
\begin{equation}\label{mADMbyInt}
M_{ADM} = {1 \over 32 \pi} \int_{r\ge a} 
{\hat{K}^{ij} \hat{K}_{ij}\over \psi^7} dv + {a \over 2} \int_0^\pi
\psi \sin \theta d\theta\ .
\end{equation}
The solid curve shows the result of substituting $\psi$, to second
order in $J$, into the integrals in Eq.\ (\ref{mADMbyInt}). The result
agrees with Eq.\ (\ref{mADMbypert}) to second order in $J$, but is
significantly smaller for large $J$. (The difference is due to terms
higher order in $J$; the expressions are identical up to second
order).  The dark points show the result of a multigrid numerical code
we wrote to solve the fully nonlinear initial value problem of general
relativity with axisymmetry, similar to that used by Choptuik and
Unruh \cite{ChUn}.

The results presented to this point are formally expansions to second
order in the dimensionless parameter $J/a^2$, but $a$ itself is not a
physical parameter, so the physical meaning of this expansion is
unclear.  For that reason we now convert our result to an expansion in
the parameter $J/M^2$.  (Here, and in subsequent expressions, we drop
the ``ADM'' subscript on $M$.  The symbol $M$ will always represent
the ADM mass.)
We then consider in the expressions we gave for the conformal factor,
\begin{equation}\label{startconvert}
\psi=1+\frac{a}{R}+J^2\left(
_0\psi^{2}+\ _2\psi_2^{2}\,P_2(\cos{\theta})
\right)\ ,
\end{equation}
that the parameter $a$ is really given by $a(M,J)$. One can obtain
explicit formulas for this by considering the expression for $a(M,J)$
obtained by inverting the approximate expression
Eq.\,(\ref{mADMbypert}).  Alternatively, one can keep the implicit
dependence of $a$ on $J$, and can replace it in the last step of a
calculation, using tabulated values for $a(M,J)$ obtained from the
multigrid numerical code. In this case the ``second order part'' is
the part of $\psi$ that remains after the part zeroth order in $J/M^2$
is subtracted.

If the relation of $a$ and $J$ is taken
from Eq.\,(\ref{mADMbypert}), the explicit formulas (correct to second
order) are,
\begin{equation}
\psi=\Psi^{(0)}+(J/M^2)^2\Psi^{(2)}=\Psi^{(0)}+(J/M^2)^2\left( _0\Psi^{(2)}
+\ _2\Psi^{(2)}\,P_2(\cos\theta)\right)\ ,
\end{equation}
with 
\begin{equation}
\Psi^{(0)}=1+M/2R \end{equation}\begin{equation}
_0\Psi^{(2)}=\frac{M}{R}\frac{
1+(5M/2R))(1+M/2R)^2+(M/2R)^4-(1+M/2R)^5}
{5(1+M/2R)^5}\end{equation}\begin{equation}
_2\Psi^{(2)}=-\frac{
(M/R)^3}
{10(1+M/2R)^5}\ .
\end{equation}

To obtain the background metric in the usual Schwarzschild form we
 introduce the radial coordinate $r$ with
\begin{equation}
R = \left(r - M +\sqrt{r^2-2Mr}\right)/2\ .
\end{equation}
Since the conformal factor, up to second order,  can be written as 
\begin{equation}\label{constapsi}
 \psi^4 = \left(\Psi^{(0)}\right)^4\left[1 + 4
 \left(\frac{J}{M^2}\right)^2 {\Psi^{(2)} \over \Psi^{(0)}}\right]\ ,
\end{equation}
the 3-geometry to second order is given by
\begin{equation}\label{initBYpert}
\left(1-{2M\over r} \right)g_{r r} = \frac{g_{\theta \theta}}{r^2}=
\frac{g_{\phi \phi}}{r^2\sin^2\theta}=
 \left[1 + 4 \left(\frac{J}{M^2}\right)^2 {\Psi^{(2)} \over
 \Psi^{(0)}}\right]\ ,
\end{equation}
and, in terms of the $r$ variable,  we may write
\begin{equation}
{ _2\Psi^{(2)} \over \Psi^{(0)}} = - {M^3\over 10 r^3}\ .
\end{equation}

%%%%%%%%%%%%%
\section{Perturbative evolution of the 
Bowen--York initial data}\label{sec:ev}
%%%%%%%%%%

We adopt the notation of Regge and Wheeler\cite{ReWh} for the
separation into parities and the multipole decomposition of the
perturbations.  Because we are considering axisymmetric situations,
all multipole decompositions are given in terms of Legendre
polynomials $P_\ell(\cos\theta)$.  On the initial constant-$t$
hypersurface of the BY spacetime, we can read off the perturbations of
the 3-geometry from Eq.\ (\ref{initBYpert}). The perturbations are
purely second order even parity, and in the Regge-Wheeler notation,
are:
\begin{equation}\label{initBY}
_2H_2^{BY} = _2K^{BY} = 
4(J/M^2)^2\  _2\Psi^{(2)} /\Psi^{(0)}= -4J^2/(10Mr^3)\ , 
\;\;\;\;\;\;\;\;\; 
G^{BY}=h_1^{BY}=0\ .
\end{equation}

The first order perturbations contained in the BY spacetime are those
specified by the extrinsic curvature in Eq.\ (\ref{BYK}), which can be 
reexpressed as 
\begin{equation}
K_{r\phi}=\psi^{-2}\widehat{K}_{r \phi} 
=\psi^{-2}\left(\frac{dR}{dr}\right)\widehat{K}_{R \phi} 
={3 J \over r^2\sqrt{1-2M/r}} \sin^2{\theta}\ ,
\end{equation} 
and is identical to the extrinsic curvature given in Eq.\
(\ref{KerrK}) for the Kerr geoemtry.  The explicit form of the metric
perturbations, to first order in $J$, depends on the gauge (first
order coordinate fixing) we choose. Let us choose the coordinates to
first order so that the initial BY metric is the same as the first
order Kerr metric given by Eq.\ (\ref{appgtphi}) of the Appendix.
That is, let us choose
\begin{equation}\label{gBY1}
g^{BY}_{t\phi} =   2 {J \over r} \sin{\theta}{\partial \over
\partial\theta} P_1 
\end{equation}
to be the only nonvanishing first order initial perturbation. Since
this perturbation is purely $\ell=1$, the Einstein equations require
that there be no gauge independent time variation in this
perturbation. Let us choose, therefore, to have the perturbation in
Eq.\ (\ref{gBY1}) be the first order perturbation for all time.  This
is equivalent to the statement that we are choosing coordinates, to
first order in $J$, so that the BY and the Kerr spacetimes agree, at
all times, to first order in $J$.

In principle, the radiation in the BY spacetime would be found from
the quadrupolar part of Einstein's equations to second order in
$J/M^2$. These equations contain terms linear in the $\ell=2$ second
order metric perturbations, and quadratic in the first order
perturbations. The general structure of these equations is discussed
in Ref.\ \cite{GlNiPrPucqg}. In that reference it is shown how these
second order equations can be combined into an equation like that of
Zerilli\cite{zerilliprl,zerilliprd} for first order equations. The
second order equivalent of the Zerilli equation differs only in that
it has ``source'' terms quadratic in the first order perturbations.
Since we know the first order perturbations, for all time, for the BY
spacetime, we know the source term.  The wave equation can therefore
be solved numerically and the radiation signal found from the solution
for the wave variable at large radius.

In practice, this computation can be made much simpler. Since the
first order perturbations in the BY and the Kerr spacetimes are
identical, the source terms will be identical in the second order
Einstein equations for BY and for Kerr. We can exploit this, by
evolving only the difference between the second order BY and Kerr
perturbations, following the technique used by Cunningham {\em et
  al.}\cite{cpm}.  To do this, we take as our second order $\ell=2$
variable the Moncrief\cite{moncrief} wave variable
%\begin{equation}\label{psidef}
%  _2\chi_{ M}(r,t)=\frac{r}{3}\left[
%    _2K^{(2)}+\frac{r-2M}{2r+3M}\left\{ _2H^{(2)}_2-r\partial 
%[_2K^{(2)}]/\partial r
%    \right\} \right]
%+\frac{2}{r}(r-2M)\left(r^2\partial 
%[_2G^{(2)}]/\partial r-2\ _2h^{(2)}_1
%\right)\ ,
%\end{equation}
\begin{equation}\label{psidef}
  _2\chi_{ M}(r,t)=
{(2 r -4 r M)\over (2 r +3 M)} \left(-2 r^2 
{\partial [\,_2K^{(2)}]\over \partial r}+6 r^2 {\partial 
[\,_2G^{(2)}]\over \partial} r+2 r  \,_2H^{(2)}_2 -12
\,_2h^{(2)}_1\right) + 4 r\, _2K^{(2)}
\end{equation}
in which $_2K^{(2)},\ _2G^{(2)},\ _2h^{(2)}_1$ and $_2H^{(2)}_2$ are
second order $\ell=2$ perturbations. 
%%%%%%%%
This Moncrief wave variable has two very useful features. First,
$_2\chi_{\rm M}$ is constructed only from perturbations in the
3-geometry on the initial hypersurfaces.  Second, it is invariant
under second order coordinate transformations, that is, under
transformations of form $(x^\mu)^{New}=(x^\mu)^{Old}+\xi^\mu$, in
which $\xi^\mu$ is second order.
%%%%%%%%

We are using here the same normalization for our wave function as in
Ref.\ \cite{GlNiPrPucqg}. This normalization is formally the same as
that of Zerilli\cite{zerilliprl}, except that we expand in $P_\ell$
rather than $Y_{\ell m}$; as a result our $_2\chi_{ M}$ is related
to the variable $\widehat{K}_{2M}$ of Zerilli\cite{zerilliprl} by
$_2\chi_{ M}=\sqrt{5/4\pi}\widehat{K}_{2M}$. For a discussion of
the relationship of the Moncrief and Zerilli variables, and various
normalizations, see Ref. \cite{lp1}.  It is straightforward to
generalize the analysis to arbitrary $\ell>2$.

We use the Moncrief wave functions $_2\chi_{\rm M}^{BY}$ to describe
the Bowen York spacetime, and $_2\chi_{\rm M}^{\rm Kerr}$ for
Kerr. The initial value for $_2\chi_{\rm M}^{\rm BY}$ are taken from
Eq.\ (\ref{initBY}) and the value of $_2\chi_{\rm M}^{\rm Kerr}$,
initially and for all time, are taken from the expansion in the
Appendix, from which we have,
\begin{eqnarray}
_2{H}^{\rm Kerr}_0 & = &   - {4J^2 \over 3 M r^2(r-2M)},\label{L=2Krr1} \\
_2{H}^{\rm Kerr}_1   & = & 0 \\
_2{H}^{\rm Kerr}_2  & = &  {2J^2 \over 3M^2r^2}   \\
_2{K}^{\rm Kerr}   & = &  -{2 J^2 (4 M+r) \over 3 M^2 r^3} \\
_2{G}^{\rm Kerr} & = &  -{ J^2 (2M+r) \over 3 M^2 r^3}\ .\label{L=2Krr5}
\end{eqnarray}

We now define  a ``radiative'' Moncrief wave function by
\begin{equation}
_2\chi^{\rm rad}\equiv\ _2\chi_{\rm M}^{BY}-\ _2\chi_{\rm M}^{\rm Kerr}\ .
\end{equation}
Since the first order perturbations, and therefore the source terms,
for $_2\chi_{\rm M}^{BY}$ and for $_2\chi_{\rm M}^{\rm Kerr}$ are the
same, $_2\chi^{\rm rad}$ satisfies a homogeneous Zerilli function
\begin{equation}\label{zeq}
{\partial^2\left[_2\chi^{\rm rad}(t,r)\right] \over \partial {r^*}^2}
- {\partial^2 \left[_2\chi^{\rm rad}(t,r)\right] \over \partial t^2}
-V(r^*)\ _2\chi^{\rm rad}(t,r) =0\ ,
\label{Zer21}
\end{equation}
where
\begin{equation}
r^*\equiv r+2M \ln[r/(2M)-1]\ ,
\end{equation}
and where $V$ is the $\ell=2$ Zerilli potential
\begin{equation}\label{zpot}
V(r) = 6\left(1-2{M \over r}\right){4 r^3+4 r^2 M+6 r
M^2+3 M^3
\over r^3 (2 r+3 M)^2 }\ .
\end{equation}

A word of explanation is appropriate about the properties of
$_2\chi_{\rm M}$ under a coordinate transformation. As already stated,
$_2\chi_{\rm M}$ is invariant under a transformation in which the
coordinates change only to second order. This is why we can use Eqs.\ 
(\ref{L=2Krr1}) -- (\ref{L=2Krr5}) to evaluate $_2\chi_{\rm M}^{\rm
  Kerr}$, and Eq.\ (\ref{initBY}) for $_2\chi_{\rm M}^{BY}$, though
the second order metric perturbations used are clearly in different
gauges. The point is that we know that the second order perturbations
can be brought into the same second order gauge with second order
gauge transformations, and that this has no effect on $_2\chi_{\rm
  M}^{\rm Kerr}$ or on $_2\chi_{\rm M}^{BY}$, and hence no effect on
$_2\chi^{\rm rad}$. It should be realized that $_2\chi_{\rm M}$ is
{\em not} invariant under a first order change of coordinates.  If we
were to perform, say, a first order change in the coordinates used in
the Appendix, then the value of $_2\chi_{\rm M}^{\rm Kerr}$ we would
compute would change. It is important, therefore, that no first order
change in coordinates is needed for the Kerr expansion in the
Appendix, or the BY spacetime in Sec.\,\ref{sec:by}. They are already
in the same first order gauge.

The initial $_2\chi_{\rm M}$ for this equation is simply the known
difference between the initial forms of $_2\chi_{\rm M}^{BY}$ and
$_2\chi_{\rm M}^{\rm Kerr}$, and turns out to be
\begin{equation}\label{initrad}
_2\chi^{\rm rad}|_{t=0}=\frac{2J^2}{5M^2r^3}\ 
\left(\frac{-5r^3-7Mr^2+25M^2r+60M^3}{2r+3M}\right)\ .
\end{equation}
There are no second order perturbations to the extrinsic curvature of
a constant time slice of Kerr, or in the BY initial data. The first of
these conclusions follows from an explicit computation based on the
metric in the Appendix. One finds that the extrinsic curvature
contains only odd powers of $J$. (This conforms to the intuition that
suggests that reversing the direction of $J$ should reverse the sign
of extrinsic curvature.)  The conformally related extrinsic curvature
$\widehat{K}_{ij}$ for the BY initial data is given to all orders by
Eq.\ (\ref{BYhatK}). The second order perturbations in the conformal
factor mean that the extrinsic curvature
$\widehat{K}_{ij}=\psi^{-2}\widehat{K}_{ij}$ will contain
perturbations of odd orders in $J$, due to the perturbations of even
order contained in $\psi$.
Since there are no second order contributions to the extrinsic curvature,
of either BY or Kerr, it follows that
\begin{equation}\label{initraddot}
\left.\frac{d}{dt}\left(_2\chi^{\rm rad}\right)\right|_{t=0}=0\ .
\end{equation}

The wave equation of Eqs.\ (\ref{zeq}) -- (\ref{zpot}), with the
cauchy data of Eqs.\ (\ref{initrad}) -- (\ref{initraddot}), is simply
solved numerically for $_2\chi^{\rm rad}(t,r)$, and from the solution
we can find
\begin{equation}
_2\chi_{\rm M}^{BY}(t,r) =\ _2\chi^{\rm rad}(t,r)+\ _2\chi_{\rm
M}^{\rm Kerr}(r)\ ,
\end{equation}
where $_2\chi_{\rm M}^{\rm Kerr}(r)$ is the known, time independent,
Kerr solution.  The radiative power (see Ref.\,\cite{GlNiPrPucqg})
contained in the BY spacetime is then given by
\begin{equation}
{\rm Power}=
\frac{3}{10}\left(\frac{d}{dt}
_2\chi_{\rm M}^{BY}\right)^2
=\frac{3}{10}\left(\frac{d}{dt}
_2\chi^{\rm rad}\right)^2\ , 
\end{equation}
and this is the gravitational radiation power emitted as the BY
solution settles into its Kerr final form.
The energy radiated is the time integral of this expression.  

\pagebreak
\begin{figure}[h]
%fig2
\centerline{\psfig{figure=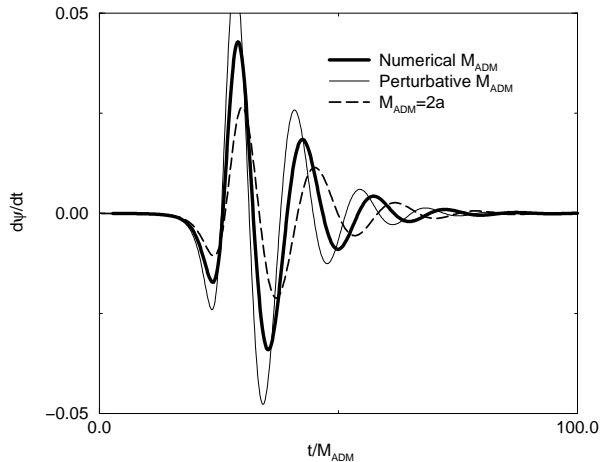,height=2.8in}}
\caption{ The radiated waveform for a single spinning Bowen and York
black hole, computed treating the spacetime as a perturbation of
Schwarzschild.  The waveform, extracted at $r=20M$, is dominated by
quasinormal ringing which persists long enough to contaminate
radiation coming out from a head on collision of black holes. We
display the waveforms corresponding to the different choices of
background mass used in perturbation theory. The significant
differences among these curves illustrates the importance of the
choice of the background mass in perturbative calculations, i.e., on
the way in which the ADM mass $M$ is taken to depend on $J$, for a
given value of $a$.  The bold curve uses the ``correct'' ADM mass
computed numerically with the multigrid code. The thin curve was
computed with expression $M=2a+J^2/(20a^3)$ (see text) which is
correct to second order in $J$. The dashed curve uses $M=2a$, and
ignores the influence of $J$ on $M$, for given $a$. Previous
experience with perturbation calculations for which numerical
relativity comparisons were available strongly suggests that the curve
with the numerically computed ADM mass is the most accurate.}
\label{fig:waves}
\end{figure}

\section{Results and discussion}

In Fig.\,\ref{fig:waves} we show the radiated waveform, as a function
of $t$, for fixed, large $r$, from which we may infer the effective
time for the decay of the Bowen--York rotating black hole into its
final Kerr state. It is clear that the wave form is dominated by
quasinormal ringing.  This means that the ``initial burst'' of energy
generated as the hole relaxes to Kerr form can contaminate the
evolution for some time.

\begin{figure}[h]
\centerline{\psfig{figure=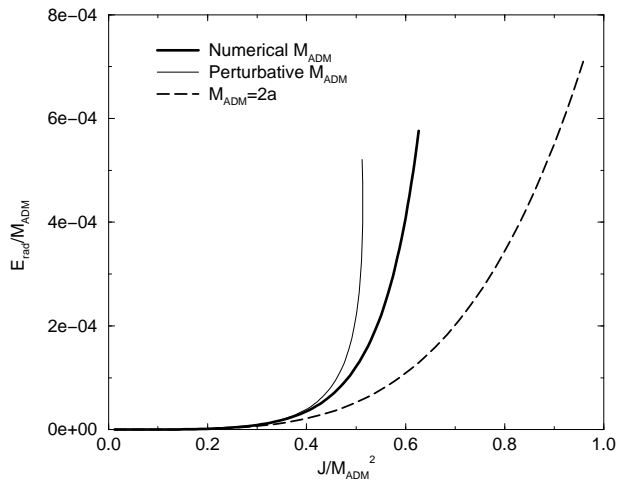,height=2.9in}}
\caption{The total radiated energy computed for ``BY relaxation.''
The method is formally correct only to lowest (fourth) order in the
expansion parameter $J/M_{ADM}^2$. For a description of the various
ADM masses involved, see the caption of figure 2.}
\label{fig:radene}\end{figure}

In Fig.\,\ref{fig:radene} we show the total energy radiated as a
function of $J/M^2$.
Whatever choice we make for the $J$ dependence of the ADM mass, our
perturbation calculation for $\chi^{\rm rad}$ is formally correct
only to second order in $J/M^2$. Since the radiated energy is
quadratic in $\chi^{\rm rad}$, the energy results displayed in
Fig.\,\ref{fig:radene} are formally correct only to fourth
order in $J/M^2$, the lowest nontrivial order.  The results in
Fig.\,\ref{fig:radene}, cannot therefore be trusted for $J/M^2$ near
the astrophysically interesting limit $J/M^2\approx1$. We suspect that
the curve corresponding to the numerical ADM mass is reliable within a
factor of two or so up to $J/M^2=0.8$. A more accurate evaluation will
require either fully nonlinear numerical relativity, or a calculation
using second order perturbations around the Kerr solution.

According to Fig.\,\ref{fig:radene}, the ``BY relaxation energy,'' the
energy emitted as a single BY hole relaxes to a Kerr hole appears to
be small. It should be kept in mind, however, that the total radiation
in a black hole coalescence can be comparably small.  For a head on
collision of nonspinning holes the total radiated energy is of order
$10^{-3}$ of the total ADM mass.  Head on collisions, of course, are
not of primary astrophysical interest. For the ``merger'' phase of
equal mass holes, radiated energy is expected to be several
percent\cite{fh}. In this case, the BY relaxation energy would be
negligibly small. It would, furthermore, be emitted within a few
quasinormal periods of a single hole, while the merger and ringdown of
the final hole formed would require a time an order of magnitude
longer.

%%%%%%%%%%%%%
\section{Acknowledgments}
We were able to write our multigrid code by studying a code of the
NCSA/Potsdam group. We wish to thank Peter Anninos for help with this.
Using the ADM mass instead of the $a$ parameter was an idea that arose
in various discussions that initially involved John Baker and Steve
Brandt.  This work was supported in part by grants NSF-INT-9512894,
NSF-PHY-9423950, NSF-PHY-9507719, by funds of the University of
C\'ordoba, the University of Utah, the Pennsylvania State University
and its Office for Minority Faculty Development, and the Eberly Family
Research Fund at Penn State. We also acknowledge support of CONICET
and CONICOR (Argentina). JP also acknowledges support from the Alfred
P. Sloan Foundation through an Alfred P. Sloan fellowship.
%%%%%%%%%%

%%%%%%%%%%%%%%%%%%%%%%%
%%%%%%%%%%%%%%%%%%%%%%%
%%%%%%%%%%%%%%%%%%%%%%%

\eject

%%%%%%%%%%%%%%%%%%%%%%%%%%%%%%%%%%%%%%%%%%%%%%%%%%
\section*{Appendix: The Kerr metric as a perturbation of a
Schwarzschild black hole}
%%%%%%%%%%%%%%%%%%%%%%%%%%%%%%%%%%%%%%%%%%%%%%%%%%

The Kerr metric in Boyer - Lindquist coordinates takes the form
\begin{eqnarray} 
ds^2 & = & {\Delta-(J/M)^2 \sin^2{\theta} \over \Sigma}dt^2
   -2 (J/M) \sin^2{\theta} {\rho^2+(J/M)^2-\Delta \over \Sigma} dt d\phi
   \nonumber \\
\label{kerr} 
&   & +{(\rho^2+(J/M)^2)^2-\Delta (J/M)^2 \sin^2{\theta} \over
 \Sigma}\sin^2{\theta} d\phi^2+ {\Sigma\over \Delta} d\rho^2+\Sigma
 d\theta^2
 \end{eqnarray} 
where $ \Sigma=\rho^2+(J/M)^2\cos^2{\theta}$ and, $\Delta=\rho^2+(J/M)^2-2 M
\rho $.

If we assume $J<M^2$, (\ref{kerr}) is defined only for $\rho > M +
\sqrt{M^2-(J/M)^2}$, since $\Delta=0$ for $\rho = M +
\sqrt{M^2-(J/M)^2}$.

The metric (\ref{kerr}) reduces to the Schwarzschild metric, in the
range $\rho > 2M$, for $J=0$. It seems reasonable, therefore, to try to
find an expansion of (\ref{kerr}), in powers of $J$, as a perturbation
of a Schwarzschild black hole. This expansion, however, would fail near
$\rho = 2M$, because the metric coefficient $g_{\rho \rho}$ does not
have the required analyticity properties. To avoid this problem we
introduce a new coordinate $r$, such that
\begin{equation} 
 r^2-2 M r=\rho^2 + (J/M)^2 - 2 M \rho
\label{Rrho}
\end{equation}
With this definition we have that $\Delta =0$ corresponds to $r=2M$.
We may  invert (\ref{Rrho}) to
\begin{equation} 
\rho = M +\sqrt{(r-M)^2- (J/M)^2}  
\end{equation}
Then, for $r> 2M$, and $J<M^2$, the right hand side may be expanded in
powers of $J$. The leading terms are
\begin{equation} 
\rho = r - {J^2\over 2M^2(r-M)}\ .  
\end{equation}
In fact, one can easily check that all the metric coefficients admit a
convergent power series expansion in $J$. To leading order we have
\begin{eqnarray}
g_{tt} & = & \left(1 - {2M \over r}\right)\left(-1 +{J^2 (r+2M) \over
3Mr^2(r-M)(r-2M)} P_0 - {4J^2 \over 3 M r^2(r-2M)} P_2 +
{\cal{O}}([J/M^2]^4) \right)\\ g_{t\phi} & = & 2 {J \over r}
\sin{\theta}{\partial \over \partial\theta} P_1 +
{\cal{O}}([J/M^2]^3)\label{appgtphi}\\ g_{rr} & = & \left(1 - {2M
\over r}\right)^{-1}\left(1 +{J^2(Mr+r^2+M^2) \over 3M^2r^2(r-M)^2}P_0
+ {2J^2 \over 3M^2r^2} P_2 + {\cal{O}}([J/M^2]^4) \right)\\ g_{\theta
\theta} & = & r^2 \left[1 - {J^2(2M^2+r^2) \over 3M^2r^3 (r-M)} P_0
-{2J^2(4M+r) \over 3 M^2r^3} P_2 -{J^2(r+2M) \over 3 M^2r^3}
{\partial^2 \over \partial\theta^2} P_2 + {\cal{O}}([J/M^2]^4)
\right]\\ g_{\phi \phi} & = & r^2\sin^2{\theta}\left[ 1 -
{J^2(2M^2+r^2) \over 3M^2r^3 (r-M)} P_0 -{2J^2(4M+r) \over 3 M^2r^3}
P_2 \right.\nonumber \\ & & \left. -{J^2(r+2M) \over 3
M^2r^3}\cot{\theta} {\partial \over \partial\theta} P_2 +
{\cal{O}}([J/M^2]^4) \right]\ ,
\end{eqnarray}
where $P_0 = 1$, $P_1= \cos{\theta}$ and, $P_2 = (3/2) \cos^2{\theta} -
1/2$ are Legendre polynomials.

From this metric it is straightforward to compute, to first order in
$J$, the extrinsic curvature of a $t=$ constant surface. If we let
$\vec{n}$ be the future directed normal to a $t=$\, constant
hypersurface, then, the extrinsic curvature is $K_{ij}=-n_{i|j}$, where
the bar denotes covariant differentiation with respect to the 3-geometry.
The normal has only a single covariant component $n_t$ which, to first
order in $J$, is
$n_t=-1/\sqrt{1-2M/r}$. With this, a straightforward computation shows that
the only nonvanishing first order components of $K_{ij}$ are
\begin{equation}\label{KerrK} 
K_{r\phi}=K_{\phi r}=\frac{3J}{r^2\sqrt{1-2M/r}}\sin^2\theta\ .
\end{equation} 

\begin{references}
\bibitem{Mi} C. Misner, Phys. Rev. {\bf D118}, 1110 (1960).

\bibitem{BoYo} J. Bowen, J. York, Phys. Rev. {\bf D21}, 2047 (1980).

\bibitem{Cook} G.~B. Cook, M. W. Choptuik, M. R. Dubal , Phys. Rev. D
{\bf 47}, 1471 (1993); G.~B. Cook, Phys. Rev. D {\bf 50}, 5025 (1994);
Ph.D. thesis, University of North Carolina at Chapel Hill, Chapel
Hill, North Carolina, 1990.

\bibitem{etaletal} P. Anninos, D. Hobill, E. Seidel, L. Smarr, W.-M.
Suen, Phys. Rev. Lett. {\bf 71}, 2851 (1993); Phys.\ Rev.\ D {\bf 52},
2044 (1995).

\bibitem{boost} J. Baker, A. Abrahams, P. Anninos, S. Brandt,
R. Price, J. Pullin, E. Seidel, Phys. Rev. {\bf D55},
829 (1997).

\bibitem{PrPu} R. Price, J. Pullin, Phys. Rev. Lett. {\bf 72}, 3297 (1994).

\bibitem{GlNiPrPucqg} R. Gleiser, O. Nicasio, R. Price, J. Pullin,
Class. Quan. Grav. {\bf 13}, L117 (1996).

\bibitem{GlNiPrPuprl} R. Gleiser, O. Nicasio, R. Price, J. Pullin,
Phys. Rev. Lett. {\bf 77}, 4483 (1996).

\bibitem{PuCa} J. Pullin, Fields Inst. Commun.
{\bf 15}, 117 (1997).



\bibitem{ChUn} M. Choptuik, W. Unruh, Gen. Rel. Grav. {\bf 18}, 813 (1986).


\bibitem{ReWh} T. Regge, J. Wheeler, Phys. Rev. {\bf 108}, 1063 (1957).

\bibitem{zerilliprl} F. J. Zerilli, Phys.\ Rev.\ Lett. {\bf 24} 737 (1970).

\bibitem{zerilliprd} F. J. Zerilli, Phys.\ Rev. {\bf D2} 2141 (1970).


\bibitem{cpm} C. Cunningham, R. Price, V. Moncrief, Astroph. J. 
{\bf 236}, 674 (1980).

\bibitem{moncrief}
V. Moncrief, Ann. Phys. (NY) {\bf 88}, 323 (1974).

\bibitem{lp1}C.O. Lousto and R.H. Price,
Phys.\ Rev.\ {\bf D55}, 2124 (1997).

\bibitem{fh}\'{E}.\'{E}. Flanagan and S.A. Hughes, preprint gr-qc
  9701039.


\end{references}
\end{document}